\documentclass[runningheads]{llncs}
\usepackage{bm}
\usepackage{graphicx}
\usepackage{cite}
\usepackage{hyperref}
\usepackage{amssymb}
\usepackage[symbol]{footmisc}

\begin{document}

\title{Geometric Transformation Uncertainty for Improving
3D Fetal Brain Pose Prediction from Freehand 2D Ultrasound Videos}
\titlerunning{Uncertainty for Improving
3D Fetal Brain Pose Prediction}

\author{J. Ramesh et al.}
\author{Jayroop Ramesh\inst{1} \and 
Nicola Dinsdale\inst{1} \and
the INTERGROWTH-21$\textsuperscript{st}$ Consortium\inst{2} \and
Pak-Hei Yeung\inst{1,3}* \and
Ana I.L. Namburete\inst{1}*}
\authorrunning{J. Ramesh et al.}

\institute{Oxford Machine Learning in NeuroImaging Lab, Department of Computer Science, University of Oxford, Oxford, United Kingdom
\and
Nuffield Department of Women's and Reproductive Health, University of Oxford, Oxford, United Kingdom
\and
School of Computer Science and Engineering, Nanyang Technological University, Singapore\\
\email{jayroop.ramesh@cs.ox.ac.uk}
}

% \institute{********}
% institute{Oxford Machine Learning in NeuroImaging Lab, Department of Computer Science, University of Oxford, Oxford, United Kingdom
% \and
% School of Computer Science and Engineering, Nanyang Technological University, Singapore\\
% \email{jayroop.ramesh@cs.ox.ac.uk}
\maketitle              % typeset the header of the contribution

\begin{abstract}
Accurately localizing two-dimensional (2D) ultrasound (US) fetal brain images in the 3D brain, using minimal computational resources, is an important task for automated US analysis of fetal growth and development. We propose an uncertainty-aware deep learning model for automated 3D plane localization in 2D fetal brain images. Specifically, a multi-head network is trained to jointly regress 3D plane pose from 2D images in terms of different geometric transformations. The model explicitly learns to predict uncertainty to allocate higher weight to inputs with low variances across different transformations to improve performance. Our proposed method, \textit{QAERTS}, demonstrates superior pose estimation accuracy than the state-of-the-art and most of the uncertainty-based approaches, leading to $9\%$ improvement on plane angle (PA) for localization accuracy, and $8\%$ on normalized cross-correlation (NCC) for sampled image quality. \textit{QAERTS} also demonstrates efficiency, containing 5$\times$ fewer parameters than ensemble-based approach, making it advantageous in resource-constrained settings. In addition, \textit{QAERTS} proves to be more robust to noise effects observed in freehand US scanning by leveraging rotational discontinuities and explicit output uncertainties.

\end{abstract}

\footnotetext[1]{These authors jointly supervised this work.}
\section{Introduction}

Two-dimensional (2D) ultrasound is the predominant tool used for routine monitoring of fetal growth, due to its cost-effectiveness, flexibility, and portability, which enable US to provide equitable obstetric examinations in low- and middle-income countries (LMICs). Optimizing the utilization of US within LMICs has the potential to reduce the number of congenital abnormalities that go undetected during early pregnancies \cite{namburete2023normative}. However, each 2D US image represents a cross-sectional view of the 3D anatomy, and localising these planes within the 3D fetal brain is difficult. Reliable and accurate localization of standard planes in fetal brain ultrasound (US) is vital to effectively analyze fetal growth and development \cite{gallery2020isuog}, and automate detection of antenatal structural defects to enable widespread deployment and adoption in limited resource settings \cite{namburete2018fully}.

Deep Neural Networks (DNNs) have been utilized to help with automated ultrasound plane detection and localization \cite{li2018standard,chen2015standard,huang2017temporal,ryou2016automated}. Despite their remarkable performance, these approaches are limited by their need for external sensor or large amounts of manually annotated data. This was addressed in \cite{yeung2021learning}, where a DNN was used to predict the 3D pose of 2D US fetal brain images using regression within a pre-defined 3D reference coordinate system. 
However, its accuracy is affected by the differences in acquisition and quality of the input images, limiting the network performance in practice. \textit{AdLocUI} \cite{yeung2022adaptive} was proposed to overcome the former limitation using unsupervised cycle consistency but does not address input variability. In this work, we aim to tackle the latter challenge using uncertainty-based approaches.
% However, this approach still suffers from poor quality pose predictions (partial recovery of fetal brain information) and abnormal localization (almost no recovery of fetal brain information). Such erroneous estimations in identifying 3D correspondence of 2D scans can lead to misjudgment of fetal brain anatomy and flawed diagnosis, and are often caused by sources including scanner differences between US device manufacturers, subjective sonographer judgment, and acquisition protocols,  \cite{yeungAdaptive3DLocalization2022}. 

We hypothesize that expanding \textit{PlaneInVol} \cite{yeung2021learning} to predict different geometric transformations representing the same 3D pose for each 2D US fetal brain image along with their respective variances can promote robustness by accounting for input-dependent noise. We thus make the following contributions:
\begin{enumerate}
    \item First, our proposed model, \textit{QAERTS}, incorporates multi-head components to regress 3D locations through diverse geometric transformations and explicitly regressed variability, inspired by ensembling \cite{lakshminarayanan2017simple}, ensemble diversity \cite{zhou2022survey}, multi-task uncertainty aware learning \cite{kendall2018multi} and geometric reprojection error \cite{kendall2017geometric}.
    
    \item Second, we adapt and extend multiple state-of-the-art uncertainty-based approaches \cite{gal2016dropout,lakshminarayanan2017simple,amini2020deep} to the task investigated in this work, namely ultrasound plane localization, as baselines for comparison.
    
    \item Thirdly, we benchmark \textit{QAERTS} and compare with baselines to empirically validate that exploiting geometric transformational variations can improve 3D localization quality compared to most baselines with only a $\sim0.1\%$ increase in parameters. Our code is publicly available at \url{https://github.com/jayrmh/QAERTS.git}.
\end{enumerate}

% \hugo{Add subtitles to each subsection to make it more organized and clearer.
% \newline
% Always relate the works you introduced to your proposed work. You should do it for each paragraph and subsection. Otherwise, it is confusing why you introduced those works.
% \newline
% If the works/methods are directly related to your proposed work or you will implement them, maybe it's better to move them to a new Background section.
% }

% During routine 2D US fetal brain scanning, one of the major goals is to acquire standard planes of view\cite{isuog_team_isuog_2021}. The biometric measurements derived from these standard planes are checked for agreement with the normal range of values corresponding to particular ages on the fetal growth chart or tables. The poor quality of these images affects the reliability of standard plane assessments and critical decisions during prenatal care.

\section{Methods}

\begin{figure}[t]
\includegraphics[width=\textwidth]{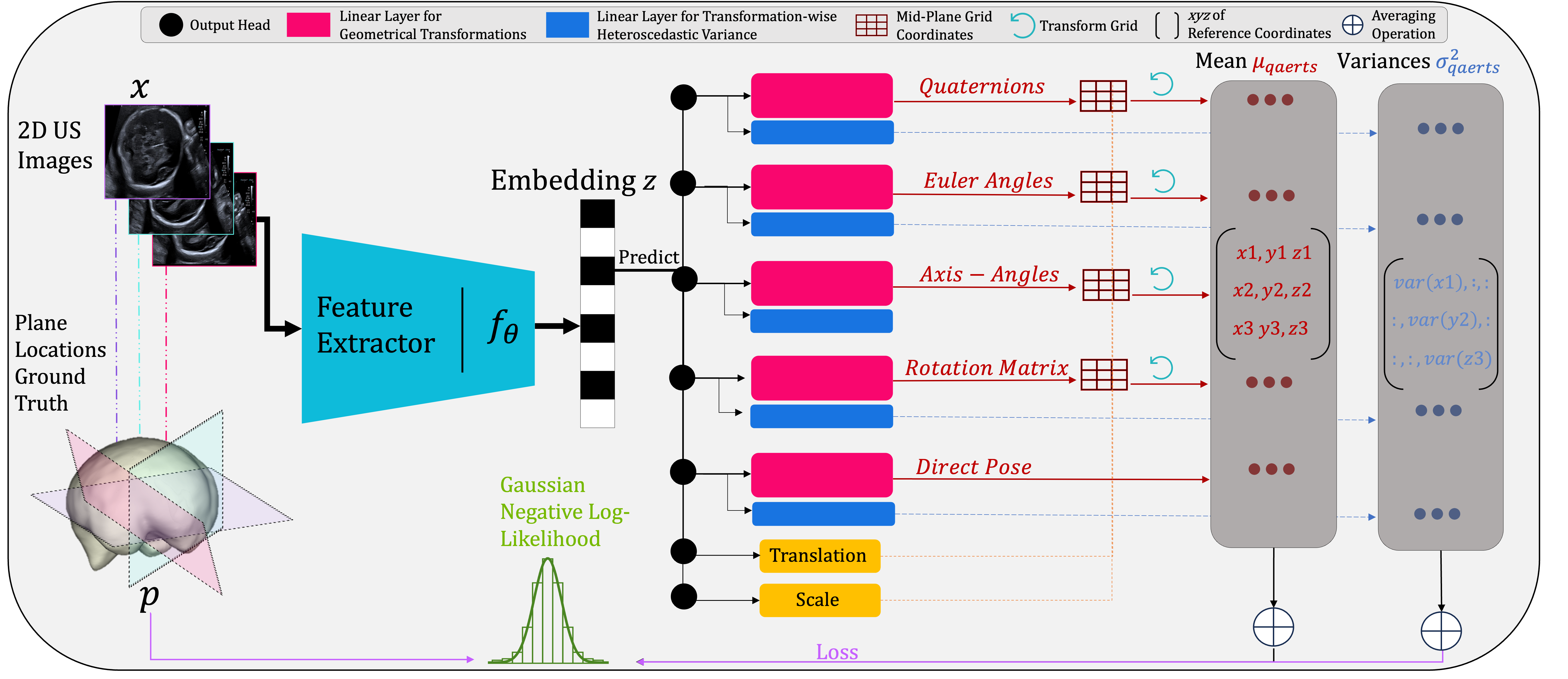}
\caption{Pipeline of our proposed work. During training, 2D slices sampled from aligned 3D volumes are augmented and used to train our proposed uncertainty-aware multi-head model with diverse parameterizations. The feature extractor (cyan) is composed of ten pairs of consecutive 2D convolutional blocks with instance normalization and rectified linear unit (ReLU) activations followed by a maxpooling operation. The generated $\bm{z}$ embedding is flattened with a adaptive pooling operation, and is propagated through two fully connected layers with a ReLU activation after each layer to a multi-head predictor (black). The trained network can be used to regress the parameters of five different geometric transformations (pink) and their resulting coordinate-wise predictive variances (blue) obtained through a set of independent fully connected layers with no activations along with shared translation and scaling parameters (yellow). The averaged 3D poses and variance obtained from an arbitrary number of 2D images is then used to compute the loss function (green).} \label{fig1}
\end{figure}

\textbf{Problem Setup:} The pipeline of our proposed work is presented in Fig.~\ref{fig1}. During training, $N$ 2D US images
$\bm{I}_{i=1}^{N}$, $\bm{I} \in \mathbb{R}^{H\times W}$ are sampled from 3D volumes aligned to the 3D anatomical atlas, meaning the true locations are known and no further human annotation is required. We adapt the prediction layer from \cite{yeung2021learning}, changing it into a multi-head network, each of which predicts different geometric transformations to express rotational quantities and direct pose predictions \cite{kendall2017geometric}. Then, we modify the network to predict learned variances in parallel such that the outputs are now Gaussian distributions, and employ maximum likelihood estimation to optimize the model weights with consideration of uncertainty.

\noindent\textbf{Model Architecture:} A feature extractor generates a fixed-length feature vector $\bm{z}$ for each input image, $\bm{I}_i$. The feature extractor $f_{\theta}(\bm{x})$ is shared across all inputs and is expected to be invariant to the permutation and number of input images. This supports our task of ultrasound image analysis as freehand image acquisition is operator-dependent and there is variability during the scanning process. The standard direct supervised pose regression model learns parameters $\bm{w}_r \in \mathbb{R}^{D\times9}$ to project the learned embedding ($D$) $\bm{z} = f_{\theta}(\bm{x})$ into a $3\times3$ matrix referring to 3 Cartesian coordinates (x, y, z) of 3 reference locations defining the predicted plane, denoted by $\hat{p}^{direct}$. We then compare with the ground truth pose defined by $p$ as follows: $\hat{p}^{direct} = \bm{w}_{r}^{T} f_{\theta}(\bm{x})$. The 3 reference locations correspond to the center, bottom right, and bottom left of the plane as per \cite{yeung2021learning}. 

\noindent\textbf{Parameterizing Rotational Representations}: The intention behind exploiting differing transformations is to separate the learning of rotational representations and variances of transformed reference coordinates, to promote diversity in the loss landscape. We refer to our proposed model as \textit{QAERTS}, where the acronym reflects the geometric transformations used: \textbf{Q}uaternions, \textbf{A}xis-angles, \textbf{E}uler angles, \textbf{R}otation matrices, \textbf{T}ranslation displacement and \textbf{S}caling. More formulation details regarding them are found in Supplementary Materials (Table 3). As shown in \cite{hempel20226d}, mapping from the $S0(3)$ space of 3D rotations to angular rotations with representations having four or fewer dimensions is discontinuous. This is because DNNs fit continuous representations with better fidelity \cite{zhou2019continuity}, and we seek to exploit the rotational discontinuities observed as a measure of the prediction ensemble uncertainty. This is likely to occur as different geometric parameterizations cover different ways of achieving the same 3D transformation and can obfuscate the parallel predictions during training \cite{salehi2018real}. Thus, we assume that the geometrical inconsistencies that can arise from utilizing the parameterizations of \textbf{A}, \textbf{E} and \textbf{R} compared to \textbf{Q} and \textit{direct pose regression} indicate an input which is \textit{harder} to learn and ambiguous. It should also be noted that large penalties will be avoided, as the loss functions are minimized with respect to the transformed set of points, and not the rotational quantities themselves. 

\noindent\textbf{Uncertainty-Aware Learning:} Our work builds upon \textit{PlaneInVol} \cite{yeung2021learning} which used the mean-squared-error (MSE) loss. However, MSE does not capture predictive uncertainty, as if the variance is uniform, the Negative Log-Likelihood (NLL) is equivalent to MSE times a constant, but if it is a learnable parameter, higher weight is given to data points with lower variance as per \cite{nix1994estimating,hullermeier2021aleatoric}. We extend the base model following \cite{lakshminarayanan2017simple}, and add multiple output heads to predict Gaussian distributions, each having separate branches to predict each geometric transformation and the respective reference coordinate variances across them. This modification leads to the Mean-Variance Estimator (MVE) variant of \textit{PlaneInVol}. Optimal weights are found by maximizing the likelihood, which is equivalent to minimizing the negative-likelihood of the predictive distributions assuming an isotopic Gaussian prior (GNLL). Instead of assuming the aleatoric uncertainty to be the same for all the data samples (homoscedastic), we explicitly model the noise considering heteroscedasticity. This is because subjective sonographer judgment and potential fetal motion relative to the probe placement \cite{yeung2022adaptive} is likely to affect each frame in 2D US scans differently, as a consequence of difference expertise levels in anatomical understanding. 

\noindent\textbf{Modifying Loss:} Using the predicted parameters for each rotational representation in orthogonal space, we transform a reference $xy$ grid with respect to the mid-plane and extract the values of the same 3 reference coordinates after the transformation through each combined translation, scaling, and rotational parameterization. Since each head learns a different set of poses and variances for each input, the ensemble of  predictions can be denoted by $\hat{p_i}$=\{$\hat{p}^{Q}_i$, $\hat{p}^{A}_i$, $\hat{p}^{E}_i$, $\hat{p}^{R}_i$, $\hat{p}^{direct}_i$\}. Then, the mean $\mu(\hat{p_i})$ of the implicit ensemble of pose predictions for every input image is computed. As the network also learns the variances for each pose prediction along each head for every independent input denoted by $\hat{s_i}$=\{$\hat{s}^{Q}_i$, $\hat{s}^{A}_i$, $\hat{s}^{E}_i$, $\hat{s}^{R}_i$, $\hat{s}^{direct}_i$\}, the resulting ensemble variance $\hat\sigma_i^2$ = $\mu({\hat{s_i})}$. Following the assumption of $\mathcal{N}(\mu_{i},\sigma_i^2)$, the GNLL is minimized wrt the ground truth $p_i$ for every input image $i$. 

The loss function for the heteroscedastic Gaussian Negative Log-Likelihood is:
\begin{equation}
 \mathcal{L}_{GNLL}=\sum_i \frac{1}{2} \log \left(\hat\sigma_i^2\right)+\frac{\left(\mu\left(\hat{p_i}\right)-p_i\right)^2}{2 \hat\sigma_i^2}
 \label{eq10}
\end{equation}
where $\sigma^2$ is the average predicted variance by the model and $\mu$ is the mean of either the explicit or implicit prediction ensembles.

\section{Experimental Setup}

\textbf{Dataset:} The 3D ultrasound fetal brain volumes (160×160×160 voxels at an isotropic resolution of 0.6$mm^3$) were obtained as part of the INTERGROWTH-21st study \cite{papageorghiou2018intergrowth}. The volumes were pre-aligned to a common reference atlas space using \cite{namburete2018fully}, and further details are available in \cite{namburete2023normative}. 

\noindent\textbf{Preprocessing:}
4 training and 2 validation 3D volumes selected at 19 gestational weeks. 

\textbf{Artificial 2D Slice Sampling:} For each 3D volume in the training epoch, a dense 2D grid along the $X$ and $Y$ axis respectively is generated in the range of -80 to 80, with each $xy$ coordinate initialized to zero. These discrete coordinates will then contain the pixel values of the 2D images, sampled from the 3D volume, and interpolation is performed where the grid does not correspond to a relevant pixel value in the volume. A third dimension representing the $z$ axis is added to localize this sampling space. Intuitively, this implies that 2D planes move along the surface normals tangential to the 3 dimensions of the volume, mimicking the ultrasound probe acquisition process. This process allows for the introduction of random variation through the sampling of infinite combinations of different 2D slices from a 3D volume. Brain segmentation masks \cite{moser2022bean} were generated to minimize learning with extracranial details from the background.

\textbf{Data Augmentations:} As the 2D images are artificially sampled in a self-supervised manner (training labels derived from the nature of the data itself), we augment sampled slices to be generalizable with freehand 2D acquired videos with random in-plane rotations. They are also subject to random rotations along the $x$ axis and $y$ axis in the range -20 to 20, the $z$ axis (along the surface normal) in the range -90 to 90, and finally random translation also along the $z$ axis between -40 and 60. Lastly, we employ random scaling in the range of 0.75 to 1.8, as well as contrast and intensity modification to introduce noise to the resolution to provide further variations to the sampling process. No normalization of pixel/voxel values were performed.

\begin{figure}[!t]
\includegraphics[width=\textwidth]{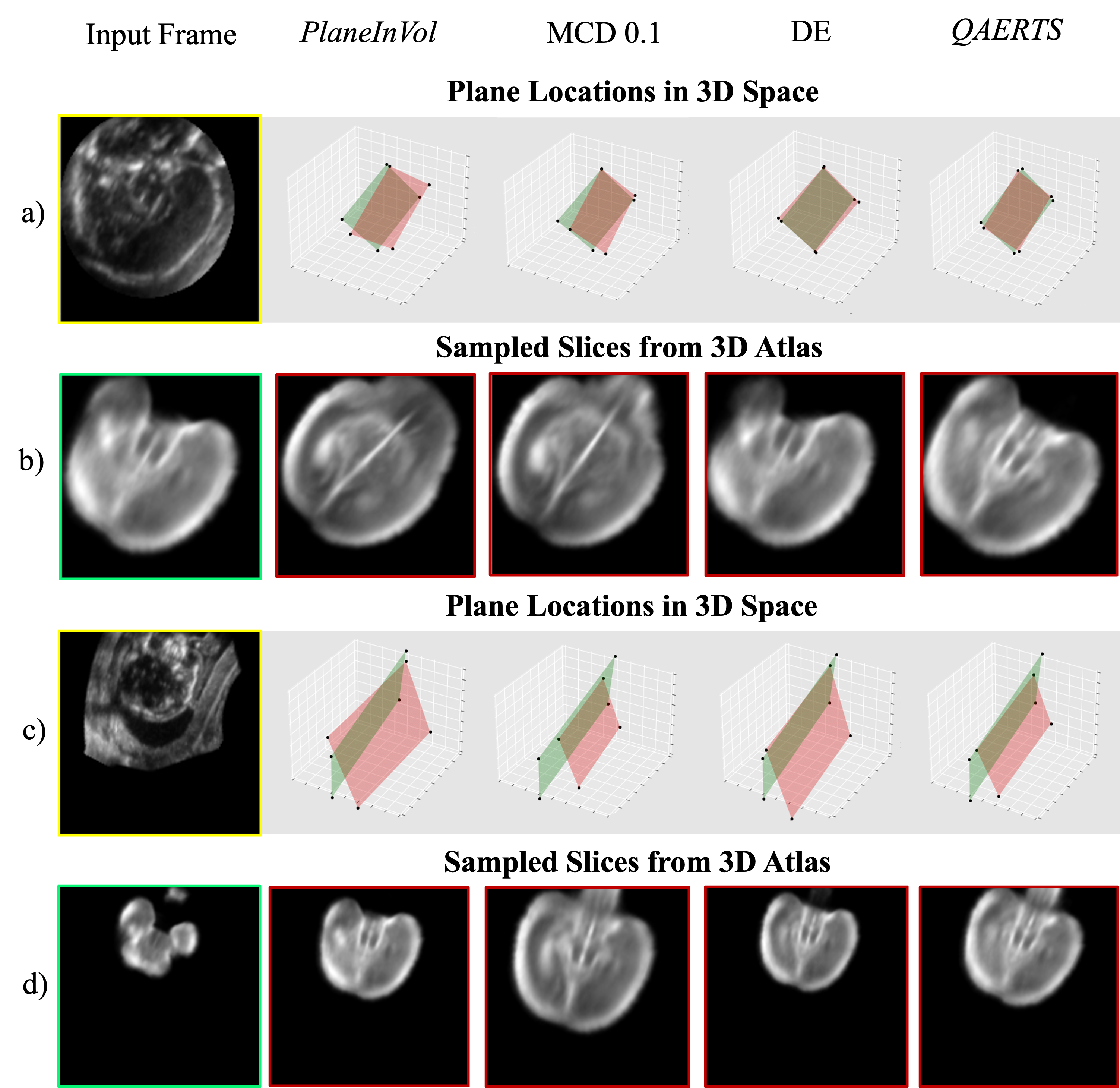}
\caption{Examples from the test set for ``high quality" and ``low quality" predictions are provided in a)-b) and c)-d) respectively.  Input frames (yellow) to the models extracted from 3D volumes are shown in the first column of a) and c). First column of a) and c) show ground-truth planes (green), and predicted (red) are visualized in 3D atlas space along the second to fifth columns of a) and c). Slices sampled (indicated by frame color) from the 3D atlas using the predicted and ground-truth plane locations for each model are shown along second to fifth columns of b) and d).} \label{fig2}
\end{figure}

\subsection{Evaluation}
\textbf{Baseline Models:} 
 We compared to the commonly deployed approaches \cite{abdar2021review} representative of each category of approach, namely: Bayesian with Monte-Carlo Dropout (MCD) \cite{gal2016dropout}, ensembling with Deep Ensembles (DE) \cite{lakshminarayanan2017simple} and deterministic methods with Evidential Deep Learning (EDL) \cite{amini2020deep} to serve as baselines against our proposed model. Specific implementation details can be found in Supplementary Materials (Table 1). All other training parameters, i.e., combinations of sampled 2D slices, epochs (5000), learning rate (0.0001), optimizer (Adam \cite{yeung2021learning}), patience (20), and batch size (49) were kept consistent between all models. All experiments utilized VGG-16 architecture with random weight initialization as a backbone.

\noindent \textbf{Testing Set}: A batch of 32 images sampled from each of the 7 3D volumes at 19 weeks of gestational age were used to quantitatively assess performance of different models. Images extracted from 2D free-hand US videos of three patients are also considered for qualitative analysis at 21 weeks of gestational age. 

\noindent \textbf{Metrics:} With the 3D anatomical atlas as a reference, we evaluate localization accuracy and sampled image quality with the following metrics. These are Euclidean distance (\textbf{ED}) and Plane angle (\textbf{PA}) between all coordinates of the predicted and ground-truth planes measured in radians, mean squared error (\textbf{MSE}) between predicted and ground truth reference points, as well as Normalized Cross-Correlation (\textbf{NCC}) and Structural Similarity (\textbf{SSIM}) between images sampled with predicted planes and ground truth planes.

\begin{table}
\caption{Mean results ($\pm$ standard deviation) for quantitative metrics during inference on the testing set. Best scores are in \textbf{bold}, and second best scores are \underline{underlined}. Upward arrows ($uparrow$) indicates higher value is better and downward arrows ($\downarrow$) indicate lower value is better. }\label{tab1}
\begin{tabular}{|c|c|c|c|c|c|c|}
\hline
Model &  ED$\downarrow$ & PA$\downarrow$ & MSE$\downarrow$ & NCC$\uparrow$ & SSIM$\uparrow$ & Parameters\\
\hline
\textit{PlaneInVol }& 0.34$\pm$0.25 & 0.42$\pm$0.25 & 277.19$\pm$245.08 &0.62$\pm$0.24 & 0.57$\pm$0.27 & $\sim$35.88M\\
\hline
MVE &0.34$\pm$0.25 &  0.40$\pm$0.25 & 224.83$\pm$241.28& 0.63$\pm$0.25& 0.60$\pm$0.25&  $\sim$35.89M\\
\hline
EDL &0.34$\pm$0.24 &0.42$\pm$0.24 & 227.90$\pm$236.49& 0.61$\pm$0.25& 0.56$\pm$0.28& $\sim$35.90M \\
\hline
MCD 0.1 &\underline{0.31$\pm$0.24}&0.45$\pm$0.25 & \underline{203.47$\pm$238.51} & 0.57$\pm$0.23 & 0.52$\pm$0.28 &  $\sim$35.89M\\
\hline
MCD 0.25 &  0.34$\pm$0.24 & 0.43$\pm$0.24 & 229.54$\pm$235.79& 0.49$\pm$0.23& 0.44$\pm$0.29&  $\sim$35.89M\\
\hline
DE &\textbf{0.29$\pm$0.23} &\underline{0.40$\pm$0.27} & \textbf{186.42$\pm$223.21} & \textbf{0.73$\pm$0.24 }& \textbf{0.63$\pm$0.24} & $\sim$141.83M \\
\hline
\textbf{QAERTS} &0.32$\pm$0.24 & \textbf{0.39$\pm$0.24}& 215.10$\pm$241.52 &  \underline{0.67$\pm$0.22} &  \underline{0.61$\pm$0.25}& $\sim$35.92M \\
\hline
\end{tabular}
\end{table}

\begin{figure}[!t]
\includegraphics[width=\textwidth]{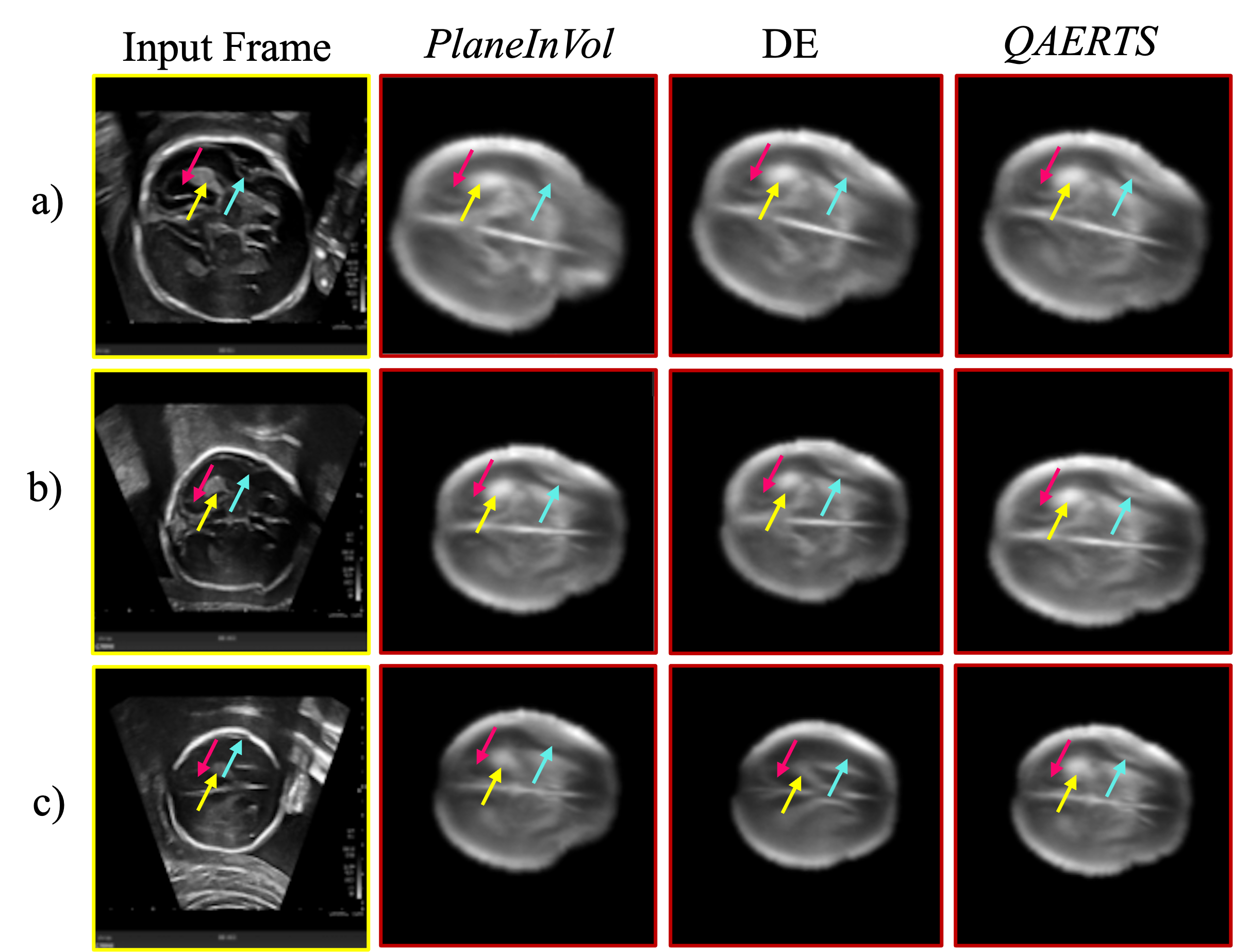}
\caption{Examples from freehand 2D US videos across three patients. The first column in a), b) and c) shows a frame from each video that were input to each model. Second to fourth column in a), b) and c) show corresponding slices sampled from the 3D atlas, using the predicted plane locations. The pink, yellow and cyan arrows indicate the anatomical structures of lateral ventricles, choroid plexus and Sylvian fissure.}\label{fig3}
\end{figure}

\section{Results and Discussion}

Relying on a single regression prediction, and not predicting a set of distributions,
leads to a larger localization error, as absolute pose regression usually generates outliers very
far from the actual scanning angles due to the instability of training regression models \cite{kendall2017geometric}. The experiments confirm our hypothesis that expanding the output layer with multiple parameterizations can reduce generalization error of the overall prediction \cite{hullermeier2021aleatoric}. 

\noindent\textbf{Quantitative Comparisons with Baselines:} From Table~\ref{tab1}, we report that the best-performing model is DE, \textit{QAERTS} and MVE across \textbf{NCC} and \textbf{SSIM} reflecting high sampled image quality. Lowest scores of \textbf{ED}, \textbf{PA} and \textbf{MSE} are once again obtained with DE, with \textit{QAERTS} and MCD in terms of plane coordinate similarity.  Without explicit ensembling, we can observe improvements of 4.457\% in \textbf{ED}, 9.268\% in \textbf{PA}, 25.225\% in \textbf{MSE}, 7.556\% in \textbf{NCC}, and 5.241\% in \textbf{SSIM} scores between the base model and \textit{QAERTS}, showing advantage of modifying the scoring criteria of the \textit{PlaneInVol} model. Interestingly, performance decreases as the dropout rate increases and both variants are among the weakest in our experiments. As DE is effectively an explicit ensembling of five MVE models, it surpasses the \textit{PlaneInVol} model significantly with \textbf{ED}, \textbf{PA}, \textbf{MSE}, \textbf{NCC}, and \textbf{SSIM} scores of 14.664\%, 6.002\%, 39.155\%, 16.470\% and 9.90\% respectively, however, this comes at a cost with $5\times$ more parameters. 

\noindent\textbf{Qualitative Comparisons with Baselines:} Fig.~\ref{fig2} show examples of ``high quality" and ``low-quality prediction" from the testing set. In practice, the former is likely when the scan contains frames concentrated around the center of the fetal brain with rich structural content, whereas the latter occurs due to partial or no recovery of fetal brain information from the peripheral regions. We use predicted locations to sample the corresponding slices from the 3D atlas to which the 3D training volumes were pre-aligned. Better prediction accuracy is obtained when the sampled slices match with the corresponding input images. From these plots, we can surmise that the additional parameterizations contributes to improved performance, and approach quality provided by explicit ensembling in the case of DE. From visualizing the predicted planes in the 3D atlas with their corresponding ground-truths show better overlap in \textit{QAERTS} when compared to \textit{PlaneInVol}, but is not as close as DE.

We assess the integrity of the models in the absence of ground-truth locations by examining the presence of anatomical landmarks in freehand 2D US in Fig.~\ref{fig3}. This is evident from the increased resolution of the lateral ventricles (pink arrow), structural fidelity of the choroid plexus (yellow arrow) and clarity of the Sylvian fissure (blue arrow) on the distal hemisphere with \textit{QAERTS} compared to \textit{PlaneInVol}. DE continues to perform better in terms of accuracy relative to the input, \textit{QAERTS} is nearly similar, and \textit{PlaneInVol} is the furthest. 

\noindent\textbf{Parameter Efficiency:} As mentioned previously, improved performance of DE comes at a computational cost, requiring approximately $5\times$ the parameters and $5\times$ the  inference latency as any other model to produce these results. MCD requires $5\times$ more inference latency to perform explicit ensembling at test time. Our proposed model \textit{QAERTS} provides considerable improvement in localization quality while maintaining parameter efficiency needing only 41,040 (0.114\%) more parameters than \textit{PlaneInVol}. Further ablation experiments are in the Supplementary Materials (Table 2).

\noindent\textbf{Loss Landscape:} A single low-loss region is likely to contain redundant solutions since the functions are almost identical \cite{kendall2017uncertainties}, and the modifications we propose in terms of different parameterizations at the output head for allow for enhanced diversity in the loss landscape relative to the single model baselines. We hypothesize this is a possible reason for performance improvement with \textit{QAERTS}, however, with DE the independently trained explicit ensemble of MVE models better estimates the true marginal likelihood \cite{lakshminarayanan2017simple}\cite{ovadia2019can}.

\textbf{Future Work:} One tangential approach to improve the proposed work is through adapting deterministic uncertainty models for our regression task of plane location prediction by incorporating distance awareness and sensitivity and smoothness in the latent space \cite{postels2021practicality} \cite{van2021feature}. A second approach would be to learn continuous representations by performing self-supervised pre-training using contrastive learning (CL) and then fine-tune for our downstream tasks \cite{assran2023self}\cite{zha2024rank}.

\section{Conclusion}

DNNs are highly parameterised, leading to a wide spectrum of possible functions that can effectively capture the data generation process. We have shown that ensemble methods lead to more consistent results, and that our proposed \textit{QAERTS} gives competitive improvement while requiring $5\times$ fewer parameters than other DE variants. Learning with geometrical parameterizations with proper scoring rules in the form of an NLL considering variance implies more retention of frames due to better pose predictions that capture more structural information. This can facilitate enhanced utility for equitable healthcare in low-resource settings by assisting sonographers in scanning guidance.

\section{Acknowledgments}
A.I.L.N. and N.K.D. are supported by the Bill and Melinda Gates Foundation.
PH.Y is funded by the EPSRC Impact Acceleration Account Award and the Presidential Postdoctoral Fellowship from the Nanyang Technological University.
%
% the environments 'definition', 'lemma', 'proposition', 'corollary',
% 'remark', and 'example' are defined in the LLNCS documentclass as well.
\newpage
\bibliographystyle{splncs04}
\bibliography{mybib}

\begin{thebibliography}{10}
\providecommand{\url}[1]{\texttt{#1}}
\providecommand{\urlprefix}{URL }
\providecommand{\doi}[1]{https://doi.org/#1}

\bibitem{abdar2021review}
Abdar, M., Pourpanah, F., Hussain, S., Rezazadegan, D., Liu, L., Ghavamzadeh, M., Fieguth, P., Cao, X., Khosravi, A., Acharya, U.R., et~al.: A review of uncertainty quantification in deep learning: Techniques, applications and challenges. Information fusion  \textbf{76},  243--297 (2021)

\bibitem{amini2020deep}
Amini, A., Schwarting, W., Soleimany, A., Rus, D.: Deep evidential regression. Advances in Neural Information Processing Systems  \textbf{33},  14927--14937 (2020)

\bibitem{assran2023self}
Assran, M., Duval, Q., Misra, I., Bojanowski, P., Vincent, P., Rabbat, M., LeCun, Y., Ballas, N.: Self-supervised learning from images with a joint-embedding predictive architecture. In: Proceedings of the IEEE/CVF Conference on Computer Vision and Pattern Recognition. pp. 15619--15629 (2023)

\bibitem{chen2015standard}
Chen, H., Ni, D., Qin, J., Li, S., Yang, X., Wang, T., Heng, P.A.: Standard plane localization in fetal ultrasound via domain transferred deep neural networks. IEEE journal of biomedical and health informatics  \textbf{19}(5),  1627--1636 (2015)

\bibitem{gal2016dropout}
Gal, Y., Ghahramani, Z.: Dropout as a bayesian approximation: Representing model uncertainty in deep learning. In: international conference on machine learning. pp. 1050--1059. PMLR (2016)

\bibitem{gallery2020isuog}
Gallery, V.V.: Isuog practice guidelines (updated): sonographic examination of the fetal central nervous system. part 1: performance of screening examination and indications for targeted neurosonography. Ultrasound Obstet Gynecol  \textbf{56},  476--484 (2020)

\bibitem{hempel20226d}
Hempel, T., Abdelrahman, A.A., Al-Hamadi, A.: 6d rotation representation for unconstrained head pose estimation. In: 2022 IEEE International Conference on Image Processing (ICIP). pp. 2496--2500. IEEE (2022)

\bibitem{huang2017temporal}
Huang, W., Bridge, C.P., Noble, J.A., Zisserman, A.: Temporal heartnet: towards human-level automatic analysis of fetal cardiac screening video. In: International Conference on Medical Image Computing and Computer-Assisted Intervention. pp. 341--349. Springer (2017)

\bibitem{hullermeier2021aleatoric}
H{\"u}llermeier, E., Waegeman, W.: Aleatoric and epistemic uncertainty in machine learning: An introduction to concepts and methods. Machine Learning  \textbf{110},  457--506 (2021)

\bibitem{kendall2017geometric}
Kendall, A., Cipolla, R.: Geometric loss functions for camera pose regression with deep learning. In: Proceedings of the IEEE conference on computer vision and pattern recognition. pp. 5974--5983 (2017)

\bibitem{kendall2017uncertainties}
Kendall, A., Gal, Y.: What uncertainties do we need in bayesian deep learning for computer vision? Advances in neural information processing systems  \textbf{30} (2017)

\bibitem{kendall2018multi}
Kendall, A., Gal, Y., Cipolla, R.: Multi-task learning using uncertainty to weigh losses for scene geometry and semantics. In: Proceedings of the IEEE conference on computer vision and pattern recognition. pp. 7482--7491 (2018)

\bibitem{lakshminarayanan2017simple}
Lakshminarayanan, B., Pritzel, A., Blundell, C.: Simple and scalable predictive uncertainty estimation using deep ensembles. Advances in neural information processing systems  \textbf{30} (2017)

\bibitem{li2018standard}
Li, Y., Khanal, B., Hou, B., Alansary, A., Cerrolaza, J.J., Sinclair, M., Matthew, J., Gupta, C., Knight, C., Kainz, B., et~al.: Standard plane detection in 3d fetal ultrasound using an iterative transformation network. In: International Conference on Medical Image Computing and Computer-Assisted Intervention. pp. 392--400. Springer (2018)

\bibitem{moser2022bean}
Moser, F., Huang, R., Papie{\.z}, B.W., Namburete, A.I., 21st Consortium, I., et~al.: Bean: Brain extraction and alignment network for 3d fetal neurosonography. NeuroImage  \textbf{258},  119341 (2022)

\bibitem{namburete2023normative}
Namburete, A.I., Papie{\.z}, B.W., Fernandes, M., Wyburd, M.K., Hesse, L.S., Moser, F.A., Ismail, L.C., Gunier, R.B., Squier, W., Ohuma, E.O., et~al.: Normative spatiotemporal fetal brain maturation with satisfactory development at 2 years. Nature  \textbf{623}(7985),  106--114 (2023)

\bibitem{namburete2018fully}
Namburete, A.I., Xie, W., Yaqub, M., Zisserman, A., Noble, J.A.: Fully-automated alignment of 3d fetal brain ultrasound to a canonical reference space using multi-task learning. Medical image analysis  \textbf{46},  1--14 (2018)

\bibitem{nix1994estimating}
Nix, D.A., Weigend, A.S.: Estimating the mean and variance of the target probability distribution. In: Proceedings of 1994 ieee international conference on neural networks (ICNN'94). vol.~1, pp. 55--60. IEEE (1994)

\bibitem{ovadia2019can}
Ovadia, Y., Fertig, E., Ren, J., Nado, Z., Sculley, D., Nowozin, S., Dillon, J., Lakshminarayanan, B., Snoek, J.: Can you trust your model's uncertainty? evaluating predictive uncertainty under dataset shift. Advances in neural information processing systems  \textbf{32} (2019)

\bibitem{papageorghiou2018intergrowth}
Papageorghiou, A.T., Kennedy, S.H., Salomon, L.J., Altman, D.G., Ohuma, E.O., Stones, W., Gravett, M.G., Barros, F.C., Victora, C., Purwar, M., et~al.: The intergrowth-21st fetal growth standards: toward the global integration of pregnancy and pediatric care. American journal of obstetrics and gynecology  \textbf{218}(2),  S630--S640 (2018)

\bibitem{postels2021practicality}
Postels, J., Segu, M., Sun, T., Sieber, L., Van~Gool, L., Yu, F., Tombari, F.: On the practicality of deterministic epistemic uncertainty. arXiv preprint arXiv:2107.00649  (2021)

\bibitem{ryou2016automated}
Ryou, H., Yaqub, M., Cavallaro, A., Roseman, F., Papageorghiou, A., Noble, J.A.: Automated 3d ultrasound biometry planes extraction for first trimester fetal assessment. In: Machine Learning in Medical Imaging: 7th International Workshop, MLMI 2016, Held in Conjunction with MICCAI 2016, Athens, Greece, October 17, 2016, Proceedings 7. pp. 196--204. Springer (2016)

\bibitem{salehi2018real}
Salehi, S.S.M., Khan, S., Erdogmus, D., Gholipour, A.: Real-time deep pose estimation with geodesic loss for image-to-template rigid registration. IEEE transactions on medical imaging  \textbf{38}(2),  470--481 (2018)

\bibitem{van2021feature}
Van~Amersfoort, J., Smith, L., Jesson, A., Key, O., Gal, Y.: On feature collapse and deep kernel learning for single forward pass uncertainty. arXiv preprint arXiv:2102.11409  (2021)

\bibitem{yeung2022adaptive}
Yeung, P.H., Aliasi, M., Haak, M., 21st Consortium, I., Xie, W., Namburete, A.I.: Adaptive 3d localization of 2d freehand ultrasound brain images. In: International Conference on Medical Image Computing and Computer-Assisted Intervention. pp. 207--217. Springer (2022)

\bibitem{yeung2021learning}
Yeung, P.H., Aliasi, M., Papageorghiou, A.T., Haak, M., Xie, W., Namburete, A.I.: Learning to map 2d ultrasound images into 3d space with minimal human annotation. Medical Image Analysis  \textbf{70},  101998 (2021)

\bibitem{zha2024rank}
Zha, K., Cao, P., Son, J., Yang, Y., Katabi, D.: Rank-n-contrast: Learning continuous representations for regression. Advances in Neural Information Processing Systems  \textbf{36} (2024)

\bibitem{zhou2022survey}
Zhou, X., Liu, H., Pourpanah, F., Zeng, T., Wang, X.: A survey on epistemic (model) uncertainty in supervised learning: Recent advances and applications. Neurocomputing  \textbf{489},  449--465 (2022)

\bibitem{zhou2019continuity}
Zhou, Y., Barnes, C., Lu, J., Yang, J., Li, H.: On the continuity of rotation representations in neural networks. In: Proceedings of the IEEE/CVF Conference on Computer Vision and Pattern Recognition. pp. 5745--5753 (2019)

\end{thebibliography}

\end{document}